\documentclass[10pt]{article}
\usepackage{epsfig,cite}
\title{Bias and temperature dependence of the
noise in a single electron transistor}
\author{Torsten Henning,
B.~Starmark, T.~Claeson, P.~Delsing\\
\textsl{Microelectronics and Nanoscience - Applied Solid State
  Physics}\\[0ex]
\textsl{G\"oteborgs Universitet / Chalmers Tekniska H\"ogskola
  AB}\\[0ex]
\textsl{SE-41296 G\"oteborg, Sweden, Fax +46~31~772-3471}}
\date{1998-10-08 (cond-mat/9810103)}
\begin{document}
\bibliographystyle{unsrt}
\maketitle
\par
\begin{abstract}
A single electron transistor 
based on Al-AlO$_x$-Nb tunnel junctions was
fabricated by shadow evaporation and in situ barrier formation. Its
output current noise was measured, using a transimpedance amplifier
setup, as a function of bias voltage, gain, and temperature, 
in the frequency range (1\dots300)\,Hz. 
The spot noise at 10\,Hz is dominated by a gain
dependent component,
indicating that the main noise contribution comes from fluctuations
at the input of the transistor. 
Deviations from ideal input charge noise behaviour
are found in the form of a bias dependence of the differential charge 
equivalent noise,
i.\,e. the derivative of current noise with respect to gain. The
temperature dependence of this effect 
could indicate that heating is
activating the noise sources, and that they are located inside or in the
near vicinity of the junctions.
\end{abstract}
\vskip3ex
\section{Introduction}

Single electron transistors with capacitive coupling ([C-]SET) 
\cite{grabert:92:sctbook,korotkov:97:cbrev}
are the
most sensitive solid state electrometers available today
\cite{korotkov:92:kochbuch}. 
They are limited in their accuracy by their noise
\cite{korotkov:94:noiseprb,krech:92:sensi,hanke:94:setapl}, 
which increases with
lower frequencies \cite{ji:94:traps}.
The empirical relation between the spectral density $S_X(f)$ of the
output quantity $X$ ($X$ being current $I$ or voltage $V$ depending on
mode of operation, or input charge equivalent $Q_\mathrm{g}$),
\begin{equation}
\label{eq:powerlaw}
S_X(f)=S_{X}(f_0)\cdot \left(\frac{f}{f_0}\right)^{-\alpha}\quad
,\alpha\approx 1
\end{equation}
has lead to the nickname ``$1/f$ noise''.
However, the deviation of
$\alpha$ from unity is often significant. We will therefore use the 
more general term
``low frequency noise''.

The low frequency noise has long since been assumed to be caused by
charged particles oscillating randomly in traps
\cite{zimmerli:92:noiseapl}, thus inducing a
displacement charge on the island, 
and shifting the operating characteristics 
of the SET by fractions of an elementary
charge. No conclusion has been reached as to the exact location of these
traps, which are generally modelled as two level fluctuators. While some
research groups expect them in the immediate (a few tenths of a
nanometre) vicinity of the island, others have seen evidence that they
might be at some distance 
\cite{zorin:96:bgnoiseprb,zimmerman:97:dist}. 
In the latter case, they would have to be in the
substrate, which is usually aluminium oxide or,
as in our case,  oxidised silicon on a silicon substrate.

A noise source in form of a charged particle trap
inside the barrier between the island and the source/drain leads,
on the other hand, might not only cause fluctuations of the island
charge, but also of the barrier's resistance. Resistance
fluctuations have been studied in larger junctions for a long time.
Such a resistance fluctuation component of the noise in 
single electron transistors has been
claimed to have been seen recently
\cite{krupenin:98:stackxxx}. 

Previously, the noise of a system consisting of a SET, its
electromagnetic environment, and the measurement setup was often
described by referring the measured (output) noise to an input charge
equivalent noise, by dividing voltage noise by the gate
capacitance or current noise by the gain. The global minimum of this
input referred noise over all bias
and gate charge values at a certain
frequency (it is customary to compare SET at 10\,Hz) was then taken as a
figure of merit, with a record low of
$7\cdot10^{-5}\,\mathrm{e\,Hz}^{-1/2}$ observed in a multilayer device
\cite{visscher:95:setapl}.

In this paper, we present extensive measurements of the low frequency
noise of a SET as a function of gate charge (or gain), bias (transport)
voltage, and temperature, in order to investigate to what extent the
noise has input noise character.

Niobium is an interesting material for single electronics, 
in comparison to aluminium prevailing to date, not only because of its
higher critical temperature and energy gap, promising increased
sensitivity of superconductive devices, but also because of the better
stability against thermal cycling and ageing from which aluminium 
devices suffer. Therefore, even the operation of niobium based devices
driven into the normal state, on which we will focus in this paper,
is of practical interest.
\section{Experimental details}
\subsection{Sample fabrication}
Although some progress has been made in the introduction of niobium as a
material for single electronics, some technological issues remain
unsolved. So far, none of the available techniques can simultaneously
fulfill all the three goals of 
\begin{enumerate}
\item high superconducting
  energy gap $\Delta$, as close as possible to the bulk value
  of 1.5\,meV;
\item as high charging energy as possible; and
\item tunnel junction resistances higher than the quantum resistance
  25.8\ldots\,k$\Omega$, but not too much higher
  than this value so as not to loose gain and,
  subsequently, output signal-to-noise ratio.
\end{enumerate}

The conventional Niemeyer-Dolan (angular or shadow evaporation)
technique 
\cite{niemeyer:74:mitt,dolan:77:masks} 
that we used produces small junctions, and
the barrier, which is formed in situ by oxidising 
the aluminium, can be
tuned to reasonable resistance values below 100\,k$\Omega$ per
junction. The price for these advantages is a rather low value of
$\Delta$ \cite{harada:94:nbset}. 
It cannot simply be explained by the low thickness of
the electrodes, which are limited to a few tens of nanometres, since
even such thin Nb films can have a $\Delta$ close to the bulk value if
deposited under more ideal conditions \cite{park:86:ultrathin}. 

The resist mask with its suspended bridges, however, prohibits the use
of surface cleaning techniques like sputtering that have been found
essential for the fabrication of high quality films
\cite{park:86:ultrathin}. In addition,
outgassing of the organic resist components due to the intense heat of
the niobium evaporation probably leads to inclusion of contaminants in
the Nb film, and the critical temperature of niobium is very sensitive
to contamination \cite{elyashar:77:apw}, especially by oxygen. 
A possible way out might be inorganic or
more heat resistant organic resists that can be pre-baked at higher
temperatures \cite{jain:85:nbpbieee}.

Other techniques  have  different drawbacks. The self-aligned
in-line technique (SAIL) gives a high $\Delta$ and low junction
capacitances, but rather high junction resistances and 
thus low current gain
\cite{bluethner:97:nbsetieee}. 
A recently published modification 
\cite{pavolotsky:98:nbtechxxx} of the
established three layer process 
using a prefabricated barrier in a sandwich
structure gives a very good $\Delta$, but it will have to be scaled down
by about half an order of magnitude in linear dimensions before the
charging energies reach those attainable by the Niemeyer-Dolan technique
at present.

\begin{figure*}\centering
\epsfig{file=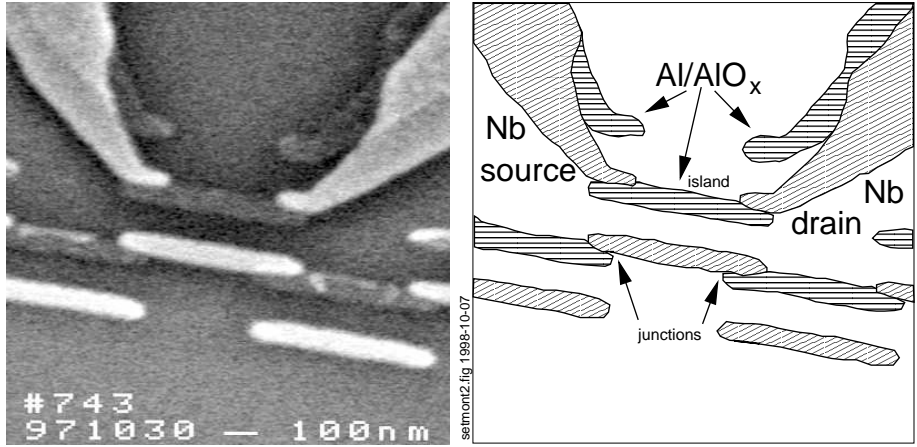,width=\textwidth}
\caption{\label{fig:setfoto}%
Scanning electron micrograph  of a single electron transistor 
nominally identical to the sample under consideration,
with Nb leads (bright) and an Al island,
and  artistic interpretation.
The excess island created by the double angle evaporation 
technique 
forms part of a linear array of junctions
not related to the measurements described here.
The gate electrode is situated outside the image area.}
\end{figure*}

Our sample substrates of size $7\times 7$\,mm$^2$  were made from
silicon wafers thermally oxidised to a depth of $(900\pm 100)$\,nm.
A gold pattern with contact leads and alignment fiducials was
produced  by photolithography. 
We used a four layer resist 
evaporation mask.
It consisted
of a bottom layer of 50\,nm 950k PMMA baked at 170$^\circ$C (to enable
liftoff), a second layer of approximately 250\,nm of Shipley S1813 photo
resist baked at 160$^\circ$C, providing support for the following layer
of 20\,nm germanium deposited by evaporation, and a top layer of 50\,nm
950k PMMA.
Electron beam lithograph patterning of the top layer was done with a
JEOL JBX 5D-II system using a 20\,pA beam at 50\,kV, the ``fifth'' lens
with a working distance of 14\,mm, and the ``first'' aperture with a
diameter of 60\,$\mu$m. The thinnest lines that were to form the SET were
designed with a width of 20\,nm, a centre-to-centre distance of 240\,nm
and an overlap of the parallel lines for leads and island of
100\,nm. Proximity correction was done manually, and the thinnest
structures, exposed at a dose of 1.7\,mC/cm$^2$, had a final line width
after processing of about 100\,nm (see fig.~\ref{fig:setfoto}).

After exposure and development for 60\,s in a 10:1 
(by volume) mixture of
isopropanol and water under ultrasonic excitation, the pattern was
transferred to the germanium layer by reactive ion etching (RIE) 
in a PlasmaTherm Batchtop VIII with
carbon tetrafluoride CF$_4$ as the process gas at a pressure of 1.3\,Pa, a
flow rate of 7.5\,$\mu$mol/s, and an RF power of 14\,W applied for
120\,s (248\,cm$^2$ electrode area, 60\,mm electrode distance). 
The layers supporting the Ge mask were then etched by RIE in the same
machine with O$_2$ as the process gas at a pressure of 13\,Pa, a flow rate
of 15\,$\mu$mol/s, and an RF power of 20\,W applied for
15\,min. These parameters gave an undercut profile sufficient for
the subsequent angular evaporation.

Evaporation of both Al 
(purity 5N) and Nb (2N8) was carried out in a UHV system with a
base pressure in the $10^{-7}$\,Pa range, equipped with a load lock for
the in situ oxidation of the barrier. First, the 
20\,nm thick Al bottom layer was
deposited, at an angle of $-21^\circ$ to the substrate normal, by
thermal evaporation from an effusion cell that delivered at a rate of
only
3\,nm/min. The resulting coarse grained structure of the Al film, with
grain sizes of tens of nanometres, would have made it impossible to
cover a Nb layer completely, as would  be necessary for 
creating the barrier in a
Nb-AlO$_x$-Nb transistor. Thus,  we chose Al as the base electrode
material. It was oxidised in non-dehumidified air at a pressure of 
8.8\,Pa for 20 min, and after pumping down for 120\,min, the Nb layer
was deposited at an angle of $+21^\circ$ to the substrate normal.
Unfortunately,  we did not carry out any pre-evaporation
of Nb against the closed shutter
for this sample. Such a procedure might have improved the quality
of the film \cite{harada:94:nbset}. 
The film was deposited by opening the shutter for
2\,s and closing it for 8\,s a total of five times. This practice was
intended to reduce damage of the mask.
Such a damage had been seen earlier, 
we had attributed it to overheating, but later found it to be caused by
fabricating the resist incorrectly. 
The interval evaporation procedure
was thus abandoned.

Two chips on a contiguous piece of substrate were processed
simultaneously. One was taken for the measurements, while the second
chip was subject to characterisation by scanning electron microscopy
(SEM). Figure~\ref{fig:setfoto} shows the image of a transistor on the
second chip corresponding to the one on the first chip on which the
measurements described in the following were performed.

\subsection{DC characterisation}The characterisation at very low
temperatures as well as the noise measurements described in
section~\ref{sec:results} were carried out with the sample attached to
the mixing chamber of an Oxford TLE~200 dilution refrigerator, reaching
a base temperature of $(30\pm 5)$\,mK. All measurement leads were
filtered by 500\,mm of Thermocoax cabling 
\cite{zorin:95:thermocoax}. The amplifier
electronics were battery powered, and the data were read out with
digital multimeters connected through shielded room feedthrough
filters. 
\subsubsection{Normal conducting state}
The sample's resistance, i.\,e. the combined resistance of both
junctions $R_\mathrm{T}=R_1+R_2$, was measured between room temperature
and 4.2\,K  one week after fabrication. It rose from $(125\pm
5)$k$\Omega$ at room temperature to $(165\pm 8)$k$\Omega$ 
at 4.2\,K
and high bias. 
The differential resistance around zero bias increased to about
215\,k$\Omega$ due to the Coulomb blockade.

We did not find any significant change of $R_\mathrm{T}$ between this first
characterisation and the 
subsequent  characterisation and
noise measurements at very low temperature, that were started one and
three months after fabrication, respectively (all data presented here
stem from the second set of measurements).

\begin{figure*}
\centering
\epsfig{file=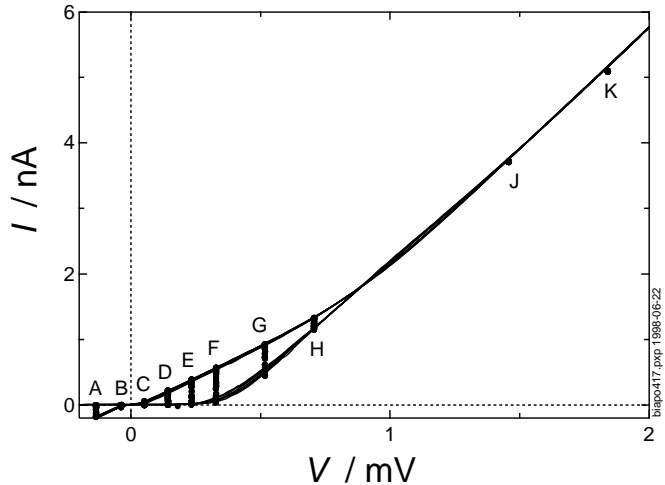,width=0.8\textwidth}    
\caption{\label{fig:biapoinv}%
$I$-$V$ characteristics of the single electron transistor 
at base temperature in the normal conducting state,
with maximum and minimum blockade. The letters indicate the voltage bias
points for the noise measurements.}   
\end{figure*}

Figure~\ref{fig:biapoinv} shows the current-voltage characteristics
(IVC) of the sample at the dilution refrigerator's base temperature, 
when it was driven
into the normal state by an external magnetic field of 5\,T. The absence
of a Coulomb staircase in the blockade indicates that the two junctions
were fairly similar. 
Also, the spread in $R_\mathrm{T}$ values was less
than 20\,\% among four nominally identical double junctions on the same
chip. 

The island capacitance $C_\Sigma$, i.\,e. the sum of the two junction
capacitances $C_{1,2}$ and the capacitance to ground 
and gate $C_0$ (which is
negligible), was determined by an offset voltage analysis
\cite{wahlgren:95:prb} at base temperature. From an
extrapolated zero bias offset voltage $V_{\mathrm{off},0}=(325\pm
15)$\,$\mu$V, we found $C_\Sigma=(0.49\pm 0.02)$\,fF.

The gate capacitance $C_\mathrm{g}$ was determined from the periodicity
of the current-gate voltage characteristics, taken during the noise
measurements,
$V_\mathrm{p}=\mathrm{e}/C_\mathrm{g}=(0.512\pm 0.008)$\,V, giving 
$C_\mathrm{g}=(0.313\pm 0.003)$\,aF.
 
\subsubsection{Superconducting case}
We found a separation in voltage between the origin and the conductance
peak at minimum Coulomb blockade in the differentiated current-voltage
characteristics of $(850\pm 20)\,\mu$V.
Using
$\Delta_\mathrm{Al}=(190\pm 10)\,\mu$eV, this means
$\Delta_\mathrm{Nb}=(235\pm 15)\,\mu$eV, or a gap 
in the niobium leads (only) 25\,\% higher
than that of 
the aluminium
island, corresponding to a critical temperature below
2\,K. We believe that 
the niobium gap should be at least twice that of aluminium
with our technique under optimised deposition conditions.
\subsection{Noise measurement setup}
\begin{figure*}
\centering
\epsfig{file=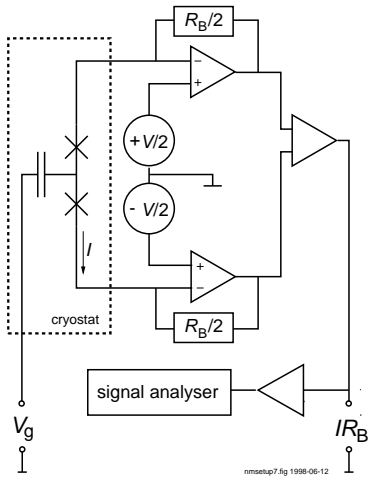,width=0.4\textwidth}
\caption{\label{fig:nmsetup}%
Transimpedance amplifier setup for current noise spectrum measurements. 
The sample is voltage biased symmetrically with respect to ground, and
the amplified current signal is read out by a spectrum analyser.}
\end{figure*}

We used a
transimpedance amplifier,  described in detail earlier
\cite{starmark:97:isec}, 
for the measurements of the low frequency noise.
It  is sketched in fig.~\ref{fig:nmsetup}.
The sample was voltage biased symmetrically with respect to ground via
two operational amplifiers with feedback resistors
$R_\mathrm{B}/2=10\,$M$\Omega$. The current signal was read out by a
HP\,35565 dynamic signal analyser that performed a real time Fast
Fourier Transform of the signal. To increase resolution, the frequency
range was divided into subranges;
25 spectra were taken and averaged in a subrange ending at
100\,Hz, and 100 spectra in the next subrange evaluated up to
300\,Hz. Each 
measurement, for one combination of bias voltage and gate charge, took
approximately five minutes.

At each bias point, 21 different gate voltages were applied, covering a
range of about one and a half elementary charges induced on the
gate. The bias points are
shown superimposed on the current-voltage characteristics in
fig.~\ref{fig:biapoinv}. 
\section{Results}\label{sec:results}
\subsection{Noise spectral density}
Over the frequency range between 1\,Hz and 300\,Hz, where we made our
measurements, the amplifier noise
\begin{equation}
i_\mathrm{n,ampl}=\sqrt{\,2k_B\frac{T_\mathrm{B}}{R_\mathrm{B}}+
\frac{e_\mathrm{n}^2(f)}{2r_0^2}}
\end{equation}
(where $e_\mathrm{n}$ is the input equivalent noise of the amplifier and
$r_0=dV/dI$ the output impedance of the SET)
was almost entirely due to the thermal noise of the feedback resistors
$R_\mathrm{B}$, situated at room temperature $T_\mathrm{B}$, 
so that we assumed
$i_\mathrm{n,ampl}=(28\pm 2)\,\mathrm{fA}/\sqrt\mathrm{Hz}$ over the
whole range.

The transistor's gain $dI/dQ_\mathrm{g}$ was calculated from the gate
capacitance $C_\mathrm{g}=Q_\mathrm{g}/V_\mathrm{g}$ and the
transconductance $dI/dV_\mathrm{g}$, which 
in turn was calculated by numerically
differentiating the current and gate voltage data taken simultaneously
with the noise spectra. The sparseness in gate voltage points caused the
uncertainty in our gain determination.

Attempts to measure the gain directly by superimposing a small
ac component on the gate voltage and reading the corresponding ac
component of the current with a lock-in technique were unsuccessful.
Harmonics, subharmonics and beat frequencies, induced
by crosstalk between input and output leads,
blurred the noise spectrum 
if the ac amplitude was chosen sufficiently
high to deliver a usable output signal, given our low gate capacitance.

\begin{figure*}
\centering
\epsfig{file=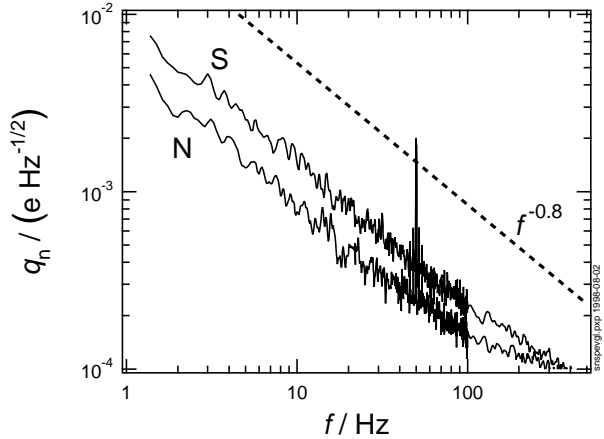,width=0.7\textwidth}
\caption{\label{fig:snspevgl}%
Charge equivalent noise 
spectra, at the bias and gate
voltage points giving maximum gain, for the normal conducting 
(\textsf{N}, cf. 
fig.\,\ref{fig:biapoinv}, point \textsf{F}) and superconducting 
(\textsf{S}) states,
respectively. The dashed line indicates the frequency dependence
$i_\mathrm{n}\propto f^{-0.8}$.}
\end{figure*}

Figure~\ref{fig:snspevgl} shows the noise spectra 
at the bias and gate voltage values with highest gain
for
both the normal and the superconducting states. Both spectra have been
referred to the input by dividing with the respective gains,
approximately 1.7\,nA/e in the normal and 3.4\,nA/e in the
superconducting case:
$q_\mathrm{n}=i_\mathrm{n}/(dI/dQ_\mathrm{g})$.
We see that the
frequency dependence
of the noise is the same in both the superconducting and the
normal states, with an exponent of $-0.8$ in the charge noise
(corresponding to $\alpha=1.6$ in the power spectrum,
eq.~(\ref{eq:powerlaw})). 
This behaviour, indicated by the dashed line in fig.~\ref{fig:snspevgl},
has also been found in all-aluminium SET on thermally oxidised silicon
substrates \cite{starmark:98:epl}.

The input charge equivalent noise in the superconducting state is
almost equal to
that in the normal state, which indicates the
input character of the noise in this frequency range 
\cite{starmark:98:epl}. At
the upper end of the frequency range, we see the crossover from input
dominated to output dominated noise.
Above 300\,Hz, the input referred noise in the superconducting state falls
significantly below that in the normal state, since the same output
current noise in both states is divided by the higher gain in the
superconducting state.

In the following, we will concentrate on the spot noise at the frequency
10\,Hz, evaluated by a
linear fit procedure in the bilogarithmic
noise-frequency diagram. We will consider the net current noise, that
is the measured current noise from which the (flat) amplifier noise and
the shot noise contribution have been subtracted. Since the shot noise
is only significant for the highest bias points well above the blockade
(contributing with 40\,fA/$\sqrt\mathrm{Hz}$ at point \textsf{K}), we
can neglect the suppression of the shot noise below and near the
blockade and use the Poisson limit
$i_\mathrm{n,Poi}=\sqrt{2\mathrm{e}I}$ \cite{korotkov:92:kochbuch}.
\subsection{Gain dependence of the current noise}
\begin{figure*}
\centering
\epsfig{file=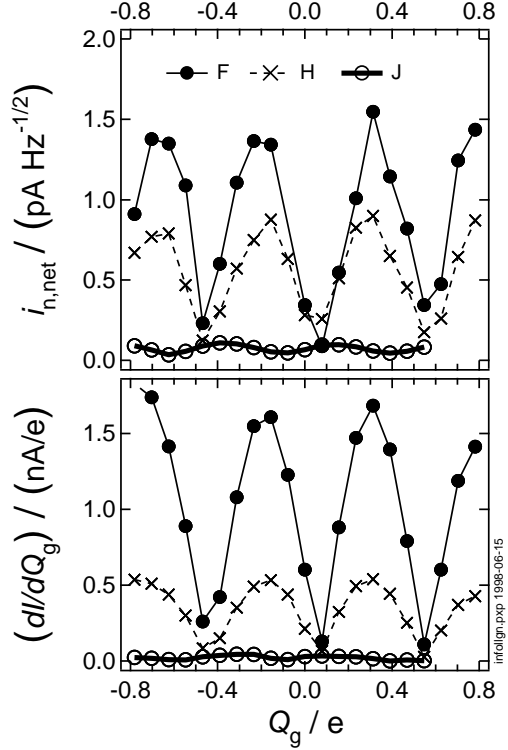,width=0.6\textwidth}
\caption{\label{fig:infollgn}%
Change of the net current noise at 10\,Hz (top panel;
amplifier noise and shot noise have
been subtracted) and of the gain (bottom panel) with the 
charge induced
on the gate of the SET. 
Bias points are labelled as in fig.\,\ref{fig:biapoinv}.
Base temperature, normal state.}
\end{figure*}
\begin{figure*}
\centering
\epsfig{file=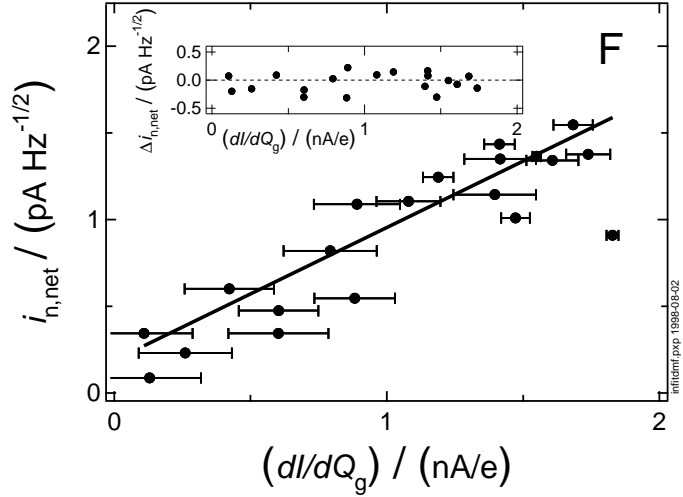,width=0.8\textwidth}
\caption{\label{fig:infitdmf}%
Dependence of the net current noise 
at 10\,Hz on the gain.
The thick line
shows is a linear least squares fit to the data points,
whose residuals are shown in the inset. The error margin on the gain
is relatively large due to the small number of gate voltage points per
bias point. Base temperature, normal state, bias point \textsf{F}.}
\end{figure*}

It is immediately evident from fig.~\ref{fig:infollgn}, showing net
current noise and gain, respectively, plotted against gate charge for
three bias points, that the noise follows the gain, or in other words,
that the noise output can in first approximation be described as charge
noise acting at the input of the SET \cite{starmark:98:epl}.
For a more quantitative
analysis, we plotted net current noise against gain
for all bias points. An example for one bias point is shown in
fig.~\ref{fig:infitdmf}.

The relation was always well described by a linear dependence, shown as
a straight line in fig.~\ref{fig:infitdmf}. We will refer to the slope
of the fit curve
$q_\mathrm{n}^\mathrm{fit}=\left\langle 
di_\mathrm{n}/d(dI/dQ_\mathrm{g})\right\rangle$, 
which has the dimension of a charge
noise, as differential charge equivalent noise.

Any deviation from pure input noise behaviour should manifest itself in
a systematic deviation from the linear relation. As we see from the
inset in fig.~\ref{fig:infitdmf}, the fit residuals are
spread fairly
randomly, so within our measurement accuracy, we cannot identify another
noise component
with gain dependence, like the correlation between resistance noise
and charge noise. 
For this correlation noise,
the square of the current noise, $S_I=i_\mathrm{n}^2$, should depend
linearly on the gain \cite{starmark:98:epl}.

The second order input charge noise contribution can generally be
described \cite{starmark:98:epl} by the coefficient $\alpha$ 
in the expansion
\begin{equation}
S_{I_Q}(f)\approx\left(\left(\frac{\partial I}{\partial Q_\mathrm{g}}
\right)^2+\frac{\alpha}{4}\left(\mathrm{e}\frac{\partial^2 I}{\partial
   Q_\mathrm{g}^2}\right)^2\right) 
S_{Q_\mathrm{g}}(f),
\end{equation} 
where $S_{I_Q}(f)$ is the output noise generated by the input charge
noise $S_{Q_\mathrm{g}}(f)$, and $\alpha$ can be evaluated as
\begin{equation}
\alpha(f)=\frac{1}{\mathrm{e}^2 S_{Q_\mathrm{g}}(f)}
\int_{-\infty}^{+\infty} S_{Q_\mathrm{g}}
(f^\prime)S_{Q_\mathrm{g}}(f-f^\prime)\,df^\prime.
\end{equation}
We found that $\alpha\approx 10^{-4}$, practically independent of
frequency. Second order contributions from this term can thus be
neglected within our measurement accuracy.

\subsection{Deviations from ideal charge noise behaviour}
We will now inspect
closer the gain dependent noise component to see if
it behaves as we would expect for a pure input charge noise.
\subsubsection{Bias dependence}
\begin{figure*}
\centering
\epsfig{file=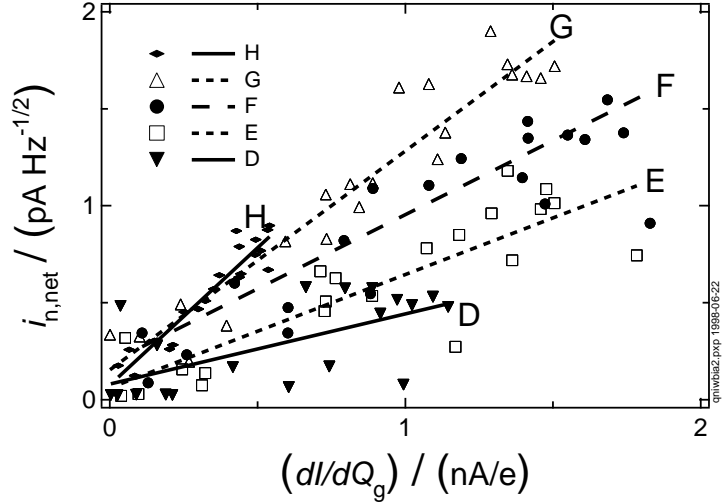,width=0.8\textwidth}
\caption{\label{fig:qniwbia2}%
Gain dependence of the net current noise (amplifier noise and shot noise
have been subtracted) at 10\,Hz (base temperature,
normal state).
With increasing bias
(\textsf{D}\dots\textsf{H}, cf. fig\,\ref{fig:biapoinv}), the ratio between
noise and gain increases, from $0.36\times
10^{-3}\,\mathrm{e}/\sqrt\mathrm{Hz}$ at bias point 
\textsf{D} to
$1.42\times 10^{-3}\,\mathrm{e}/\sqrt\mathrm{Hz}$ at point 
\textsf{H}. 
These slopes have been determined by a least square fit to the data
as illustrated in fig.\,\ref{fig:infitdmf},
error bars and residuals have been omitted to reduce
clutter.}
\end{figure*}

In
fig.~\ref{fig:qniwbia2}, the linear  fit of current noise versus
gain relation
from fig.~\ref{fig:infitdmf} is shown for the five bias points
around the global gain maximum. Comparing with the nomenclature of
fig.~\ref{fig:biapoinv}, it is obvious that the slope of the fit curves,
the
differential charge 
equivalent noise, increases 
with the bias voltage.

In the simple model of low frequency noise in SET
\cite{starmark:98:epl}, 
we would expect such a dependence only as a second order effect, via
a bias dependence of the  input charge noise itself.
The observed bias dependence
indicates that the noise sources must be located quite close to the
current path, since it seems implausible that distant defects should be
affected by the small transport voltages or currents.
The immediate practical implication of the bias dependence of the
output noise is that for low noise
operation, a SET should be operated in the low bias region, where of
course a tradeoff against signal amplitude will have to be made.
\subsubsection{Temperature dependence}
A possible mechanism, via which the bias could influence the noise
sources, is heating of the barriers, the island, and the leads and
surfaces in their vicinity, by the  dissipation near the
junctions. This was suggested as an explanation of the 
observed weak current dependence ($\propto I^{1/4}$)
of the low frequency noise \cite{wolf:97:cpem}. Measurements of the
temperature dependent behaviour of a single two level fluctuator
\cite{kenyon:98:tlftempprepr} corroborate this explanation, if one
agrees with the common assumption that the low frequency noise is the
effect of a large number of uncorrelated such two level fluctuators.

\begin{figure*}
\centering
\epsfig{file=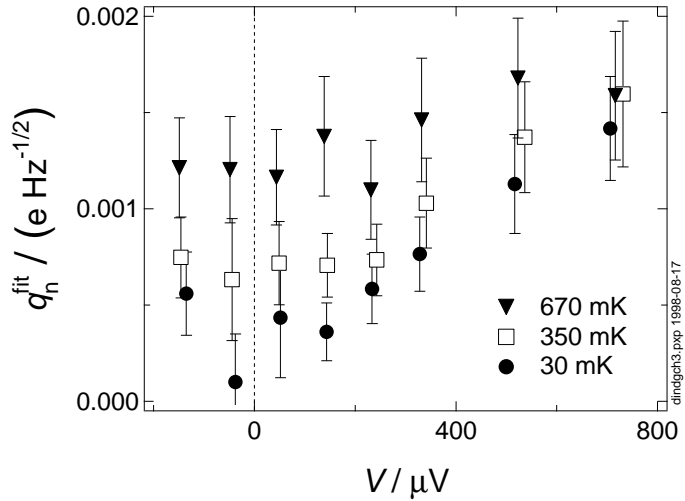,width=0.8\textwidth}
\caption{\label{fig:dindgch3}%
Differential charge equivalent noise
(proportionality constant relating gain increase to
current noise increase), as a
function of bias voltage and at different temperatures.
The values were determined as the slopes of the linear fit curves
in the noise current versus gain diagrams
(cf. fig.\,\ref{fig:infitdmf}). The error margins were estimated
from the average amplitude of the fit residuals.}
\end{figure*}

The measurements at base temperature, described above, were
repeated at temperatures of 350\,mK and 670\,mK to test the
heating hypothesis. Simple model calculations let us expect a
self-heating of the SET to about half a Kelvin at the upper end of our
bias range.
Figure~\ref{fig:dindgch3} shows the differential charge equivalent
noise, calculated by the procedure demonstrated in
figs.~\ref{fig:infitdmf} and \ref{fig:qniwbia2}, plotted against bias
voltage for the three temperatures. The error margin has been estimated
from the average amplitude of the fit residual.

For base temperature, we see the clear increase of the differential
charge equivalent noise with bias voltage that we found earlier. With
increasing temperature, the zero bias value becomes significantly
different from zero, and at a temperature of the order of 
half a Kelvin, the bias voltage dependence has vanished.

At the highest bias points (\textsf{J} and \textsf{K}), 
the differential
charge equivalent noise was masked completely by zero gain noise and
could not be determined.

This temperature dependence partly lifts the 
above stated possibility of minimising the  noise by
operating the SET at low bias.
At higher temperatures, the input noise becomes independent of the bias,
and therefore one can only minimise the noise in the SET by maximising
the gain.
\subsubsection{Zero gain noise}
Another deviation from the ideal input charge noise behaviour is the
gain independent component that we call ``zero gain noise'' or ``excess
noise'' \cite{starmark:98:epl}. It can simply be determined as the
offset along the
noise axis  in the fit procedure used for calculating the
differential charge equivalent noise.

\begin{figure*}
\centering
\epsfig{file=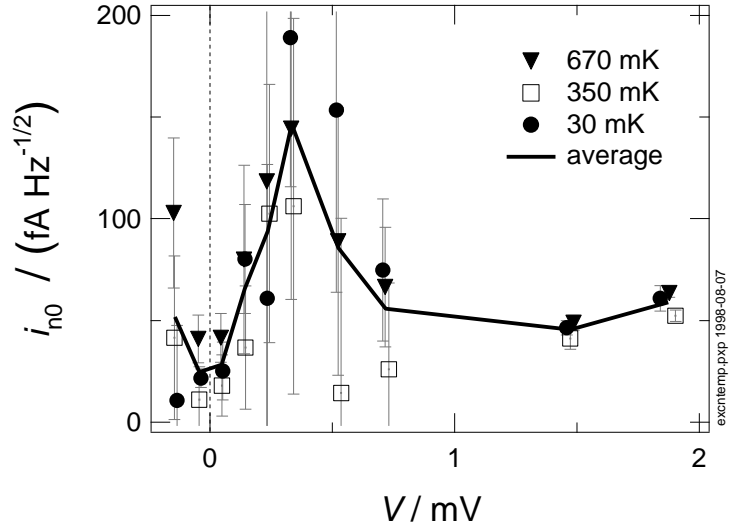,width=0.8\textwidth}
\caption{\label{fig:excntemp}%
Zero gain noise at 10\,Hz (normal conducting state)
as a function of bias
voltage,
for the same temperatures as in fig.~\ref{fig:dindgch3}. 
The values were calculated from the vertical axis intersection
in the fit procedure illustrated in fig.~\ref{fig:infitdmf}, the errors have
been estimated from the ratio between the average amplitude of the fit
residuals and the gain range.}
\end{figure*}
Figure~\ref{fig:excntemp} shows the zero gain noise as a function of
bias. The dependence is essentially the same as that of the excess noise
calculated earlier as an integral in the frequency band between 50 and
100\,Hz \cite{starmark:98:epl}. In any case, the zero gain noise has a
peak around the bias point where the current modulation is maximal, and
is practically independent of temperature. At the present time, we have
no conclusive interpretation of the cause of this excess noise.
\section{Conclusions}
In studying the low frequency noise of a single electron transistor, we
found that the output current noise is dominated by a component
proportional to the gain of the transistor, which can be described as
input charge noise. 
We found that the noise level of the transistor, expressed as the
coefficient relating output noise to gain, increases with the bias
voltage.
At low temperature, low bias conditions are preferrable for low noise
operation of the SET. The bias dependence 
of the noise 
could indicate that the current through the SET is activating the
background charges.
This
could be in turn be interpreted as a
heating effect, corroborating the
general belief that the charge noise sources are situated inside or in
the near vicinity of the tunnel junctions. At higher temperature, the
bias dependence of the noise disappears, and the transistor should be
operated at maximum gain for optimal noise properties.
\section*{Acknowledgements}
Discussions with A. N. Korotkov on modelling the low frequency noise
in SET
are gratefully acknowledged.
Our samples were made using the Swedish Nanometre Laboratory,
G\"oteborg. This work is part of the ESPRIT project 
22953 CHARGE, and we
were supported by Stiftelsen f\"or Strategisk Forskning as well
as the Swedish agencies NFR and TFR.
\bibliography{tynoise} 
\end{document}